\documentclass[runningheads,12pt]{llncsx2}
\usepackage{epsfig}

\usepackage[]{endnotes}

\usepackage[]{eurosym}

\let\footnote=\endnote

\def\tm{\raise5pt\hbox{{\rm\tiny TM~}}}

\usepackage{hyperref}

\begin{document}

\pagestyle{headings}
%In order to omit page numbers and running heads
%please change this line to
%\pagestyle{empty}
%and change the first command line too, see above.

\mainmatter

\title{Open Access, library and publisher competition, \\
and the evolution of general commerce}

% \title{The positive effects of publisher 'Big Deals' \\
% and other reflections on the evolution \\
% of scholarly communications and general commerce}

% \title{In praise of academic publishers' 'Big Deals' \\
% and other heretical reflections on the evolution \\
% of scholarly communications and general commerce}

\titlerunning{Scholarly publishing and evolution of commerce}

\author{Andrew Odlyzko}

\authorrunning{Andrew Odlyzko}

\institute{School of Mathematics\\
University of Minnesota\\
% 127 Vincent Hall, 206 Church St. SE\\
Minneapolis, MN 55455, USA\\
\email{odlyzko@umn.edu}\\
% \texttt{http://www.dtc.umn.edu/$\sim$odlyzko}}
\texttt{http://www.dtc.umn.edu/$\sim$odlyzko}\\
\texttt{Preliminary version, February 4, 2013}
}

\maketitle

\begin{abstract}

Discussions of the economics of scholarly communication are
usually devoted to Open Access, rising journal prices,
publisher profits, and boycotts.  
That ignores what seems a much more important development
in this market.  Publishers, through the oft-reviled ``Big Deal'' packages,
are providing much greater and more egalitarian access to the
journal literature, an approximation to true Open Access.  
In the process they are also marginalizing libraries, and
obtaining a greater share of the resources going into 
scholarly communication.  This is enabling a continuation
of publisher profits as well as of what for decades has been called ``unsustainable journal price
escalation.''  It is also inhibiting the spread of Open Access,
and potentially leading to an oligopoly of publishers controlling
distribution through large-scale licensing.
% regional or national licensing.

% ~~~~Whether publishers can maintain their profitability by
% squeezing more out of libraries is an open question.
% But in any case, the
~~~~ The ``Big Deal'' practices are worth studying for
several general reasons.  The degree to which publishers succeed
in diminishing the role of libraries may be an indicator of the
degree and speed at which universities transform themselves.
More importantly, these ``Big Deals'' appear
to point the way to the future of the whole economy, 
where progress is characterized by declining privacy,
increasing price discrimination, increasing opaqueness in
pricing, increasing reliance on low-paid or upaid work
of others for profits, and business models that depend on
customer inertia.

\end{abstract}

\section{Introduction}\label{sec:introduction}

Concerns about libraries not being able to afford rapidly
escalating journal costs 
go back many decades.  Over the
last 15 to 20 years, they have intensified.  This
was partially because
the cost pressures have become more burdensome.  Perhaps even
more important has been the arrival of the Internet, as well as
of computerized typesetting and other modern electronic 
tools.  These have offered the possibility of much less expensive
methods of scholarly publishing.  As one result, there has been an
increased push for various forms of Open Access (which is also
being advocated for other reasons, in order to make information
more widely available).
There have also been calls for scholars to stop
collaborating with the expensive commercial publishers,
by refusing to submit their papers to them, as well
as to referee submissions or edit journals.  The
most recent prominent call of this nature was initiated
by Tim Gowers in early 2012, and as of this writing
has attracted over 13,000 signers~\cite{Costofknowledge}.

Protesters, such as those who endorse the boycott~\cite{Costofknowledge}, tend to
cite the 
% very 
high profits of commercial publishers, most commonly
of Elsevier, the largest one, as injurious to scholarly communication,
and unjust, being based on donated labor of academics.  They
also often complain about the ``Big Deals'' that
large publishers, again with Elsevier in the forefront,
force libraries into, cf.~\cite{ArnoldC2012,Bergstrom2010}.  
In these contracts, which are universally
shrouded in secrecy (although an interesting project
is revealing some of the details~\cite{BergstromCM}),
libraries are forced to accept multi-year commitments with
steady 
% and rapid 
price escalation and little flexibility in 
selecting what journals they get.  This has all the obvious
disadvantages for libraries and the academic community.
However, such discussions almost universally ignore the
positive effects of the ``Big Deals,'' as well as the degree
to which those positive effects are key to
the main action in scholarly publishing, namely the competition
between libraries and publishers.
%  for positions and resources.

\begin{figure}
\centerline{\epsfig{figure=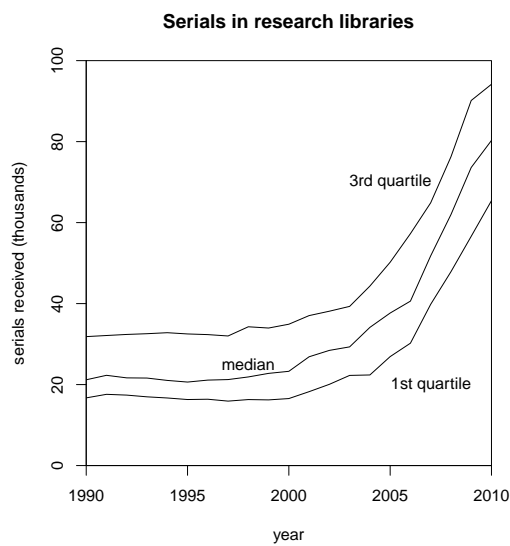,height=10.2cm,angle=-90}}
\caption{Number of serials available in research libraries from 1990 to 2010.
Shows the median as
well as the first and third quartiles of the number of serials received
by academic members of ARL.}
\label{fig:serials}
\end{figure}

The main contribution of this paper is an investigation of
the effects of the reviled ``Big Deals'' on access to scholarly journals.
In spite of their low reputation, over the last decade they
have produced a remarkable increase in availability of serials.
% (They have also surely retarded the spread of Open Access, a negative influence to be discussed later.)
This is shown in~Fig.~\ref{fig:serials} and is discussed in
more detail in Section~\ref{sec:bigdeals}.
The data for the analysis is taken from the careful and
detailed statistics for the 115 academic members
of the Association of Research Libraries (ARL) \cite{ARL}.
Those members include almost all large university
libraries in the U.S. and Canada.  (For more about ARL, its statistics,
and the selection of data, see Section~\ref{sec:arl}.)
The median of the number of serials received by ARL members almost
quadrupled during the period under investigation, 
going from 21,187 in the 1989-1990 academic year
to 80,292 in the 2009-2010 one.  Practically the entire increase took 
place during the last half a dozen years, without any big changes in
funding patterns,
and appears to be due
primarily to ``Big Deals.''

Members of ARL are all large libraries, but with substantial variation.
In the 2009-2010 academic year
their budgets (as defined by ARL) ranged from \$8.3 to \$111.6 million, with
a median of \$22.8 million.  Thus improved access to journals for just
these institutions still leaves out in the cold the general public
as well as students and researchers at many hundreds of other higher
education institutions in North America, as well as potential readers
in other countries.
(It should be noted, though, that many of those institutions
do appear to be benefiting from ``Big Deals'' and other arrangements,
it's just that this study is limited to ARL members by the nature
of the data that is was available .)
It also ignores wider issues of access to other types of information, cf.~\cite{Maxwell2012}.
Still, these 115 institutions
do contain a very substantial fraction of both authors of scholarly journal papers
as well as of readers.  Thus it is of interest to see what the recent
trends have meant for them, especially since there are analogies and
implications for the wider economy.

Fig.~\ref{fig:serials} demonstrates not only that the average number
of serials available in ARL libraries has grown, but that 
gaps between the large and small libraries have decreased.
This is shown in more detail in Section~\ref{sec:bigdeals}, 
especially in Fig.~\ref{fig:ratio}.
Section~\ref{sec:arl} proves that this occurred without
any dramatic changes in either relative or absolute spending
on serials.  This socially positive development is the outcome
of the growth in price discrimination, the selling of the same
product or service at prices varying depending on customer.  
As an example,
in 2007 unlimited access to the entire collection of journals
published by Elsevier cost the University of Michigan
\$1,961,938.75, while the University of Montana paid \$442,224.78.
Such practices are increasingly common, although
usually carefully hidden from view.  (Uncovering
this degree of differential pricing required
considerable effort by
Ted Bergstrom and his collaborators, including fighting a lawsuit
from Elsevier~\cite{BergstromCM}.)  
Hence the spread of ``Big Deals'' provides 
interesting perspectives on the development of the modern economy,
a topic pursued in the final sections of this paper.
Before considering that wide subject, let us return to the
scholarly journal crisis.

Can commercial publishers continue to prosper?  Financial
analysts are divided.  A team from Exane Paribas in Paris
declared (as cited in~\cite{PolyMath}) that the boycott initiated by Gowers~\cite{Costofknowledge}
was effectively a tempest in a teacup, and that sales of Elsevier shares by 
investors foolish enough to be scared
offered a ``trading opportunity.''
On the other hand, Claudio Aspesi and his colleagues at Bernstein Research
in London have been bearish on the commercial STM (science,
technology, and medicine) publishing sector for several years (cf.~\cite{AspesiS20090716,AspesiS20110419}
and other reports) and decided the recent events were 
foretelling serious trouble~\cite{AspesiRW20120206}.

The bearish views on Elsevier are based largely on the estimate
that the ``Big Deals'' are becoming unaffordably expensive and will soon
start breaking down.  That might happen.  However, the complaints
about unaffordable journals go back many decades,
yet they somehow have been afforded.
To understand how this was possible, one needs to take a larger view of the
economics of scholarly communication.  The high profits
earned by commercial publishers, as well as by many non-profit
professional societies, are one of the lesser inefficiencies
in that system.  Those profits are just a fraction of the
actual, and now 
% (in my considered opinion) 
unnecessary, real
costs of the current scholarly journals.  And those real
publishing costs are much smaller than the (now) unnecessary real
costs of the library system.  Publishers have managed to continue
with their ``unaffordable'' journal price increases by squeezing
some of that inefficiency out of the libraries.  And there is
a lot more to be squeezed.

Discussions of the economics of scholarly publication almost uniformly
ignore the dominant economic factor in that area, namely that most of the cost is
inside libraries\footnote{As in
all of this paper, only the large North American institutions that  
are represented in the ARL are considered.  The cost structure  
is often different in other places, in particular in the  
less-developed world, where labor costs are lower.}, 
and that technology is making feasible and
desirable the dramatic downsizing of traditional functions not just of publishers,
but also of libraries~\cite{Odlyzko1994,Odlyzko1999}.
(Those costs, in turn, are dwarfed by the costs to authors,
editors, and referees, cf. Section~\ref{sec:futurology}, but those costs
do not involve any money transfers and are not accounted for.)
As a prime example, while librarians increasingly take on new roles, handling the
huge volumes of books and bound serials is still a large part
of the cost of library systems.  Yet, at a rough estimate,
90\% of the books and 99\% of the journals in a typical large
academic library can, should, and will be sent to inexpensive
off-site warehouses, with usage shifting to electronic copies.
That will free up space and eliminate many of the jobs in the system.

% This is likely less the result of a conspiracy of silence, and more
The key role of high internal library costs in academic information
systems is probably ignored because
of a general reluctance to face unpleasant facts\footnote{One of the few
recent items that even touched on the subject was a post by Phil Davis~\cite{Davis20120215}
that cited the decreasing fraction of university budgets available for
internal expenses as a sign of greater efficiency inside libraries.
As is argued here, it is more a case of publishers performing more
of the work that was traditionally in the domain of librarians
than of librarians being more efficient.}.
Libraries are central to the image of modern universities,
the repositories of knowledge and wisdom.  They are
unifying institutions serving everybody on campus and
often in the community at large.  Their buildings are
usually centrally located and among the most imposing
around, and they are staffed by the helpful librarians, 
perhaps the best-loved group on campus.
The notion of laying off those friendly librarian colleagues and
closing down library facilities in
order to sustain the profits of Elsevier is repugnant.
But that is what we are facing.  Furthermore, sending physical collections
to inexpensive off-site warehouses, closing down
physical library facilities, and limiting hours of operation of remaining
ones, all of which has been taking place to only a limited extent so far, is
not only inevitable, but desirable.  Online access is more
effective and is bound to dominate to an increasing extent.
Such a transformation has already taken place 
5 to an even greater extent 
in corporations,
which have largely eliminated their libraries, but do
pay for access to commercial databases, including those that contain
scholarly materials.

ARL statistics, summarized in Section~\ref{sec:arl},
show that while library budgets have grown much faster
than general inflation, they have shrunk compared to
university budgets.  At the same time, the fraction
of library costs that are devoted to purchases of
books, journals, and databases has grown over the
last two decades from about a quarter to about a third
(if we consider the full costs of library systems).
This indicates that libraries are becoming less
important to universities, while publishers have
roughly maintained their position by wresting 
more of the resources from the libraries.
(There has also been growth in spending on other
scholarly communication services outside the library/publisher
area that are not captured in the ARL statistics.)

Among recent moves towards mandated open access in Britain,
the Finch report and the RCUK Open Access policy
have been criticized by many Open Access advocates
for tilting towards publishers
through an embrace of Gold Open Access (in which publishers are
paid to make the articles in their journals publicly
available).  This is seen as potentially imposing crippling
additional costs on researchers.  However, although this
was not mentioned explicitly in those reports, a very obvious
move might be to recover those extra costs not from research
funds, but from library budgets, and force cutbacks in
traditional library functions.
Thus while Aspesi et al.~\cite{AspesiRW20120919}
have seen these recommendations as a threat to publishers,
it may actually mean another victory in their competition
with libraries.

% Far more shrinking of the costs of the traditional scholarly
% communication system is possible, as will be argued
% later.  (The funds saved from that would most likely be
% used to provide improved support for other information
% services.)
% As just one example, it appears likely that
% just the money that the ARL libraries alone spend on acquisition
% of serials would in principle suffice to fund an adequate
% Open Access
% publication support for all the scholarly journals in
% the world.
% But that is just ``in principle,'' and the
% huge inertia of academia stands in the way.

Open Access is not the same as low-cost publishing.
The two can be treated as orthogonal aspects of scholarly communication.
However, the low costs made possibly by modern technology make a
break with traditional high-cost subscription publishing models far more feasible.
Andy Grove, the famous former CEO of Intel, argued that 
a 10-fold change in cost or capability
of a technology requires fundamental rethinking 
of the basic business model.
Electronics does offer the possibility of a 10-fold decrease
of basic journal publishing costs.
As one example, it is argued later in this paper that
just the money that the ARL libraries alone spend on acquisition
of serials would in principle suffice to fund an adequate
Open Access publication support for all the scholarly journals in the world.
% that modern technology makes possible
% do make the transition to Open Access far more feasible.

When I first embarked on studies of electronic publishing almost
two decades ago,
I concluded that Open Access was not just desirable, but also inevitable.
It has manifold advantages for society as a whole and for scholars.
However, the inertia of librarians and scholars has kept
them from taking advantage of the opportunities offered
by modern technologies.  On the other hand, publishers
have moved faster, and it now appears that
subscription journals may survive for
a long time, together with their associated unnecessarily high
costs and unnecessarily high profits.

The evolution of scholarly publishing, including the
changing roles of libraries and publishers, is of interest
as a bellwether of change in academia in general.
Two decades ago, the arrival of the Internet and
other electronic technologies led to prophesies
that universities were doomed~\cite{Noam1995}.
Instead, as was easy to predict~\cite{Odlyzko1997c}, they 
blossomed.  (See Section~\ref{sec:universities} for
some statistics.)  Today a new wave of ``disruptive innovation''
is threatening the traditional academic model, with a
proliferation of for-profit educational institutions
as well as free massive online courses being cited as
just the start of massive changes.  Chances are
that, as was argued in~\cite{Odlyzko1997c}, 
there will be far less of a collapse than is promised (or threatened),
and that the new techniques will be more of an addition than
a replacement for traditional approaches.
Education, just like medicine, is a growth business in a world that depends
increasingly on training and expert knowledge.  Hence this sector as
a whole is not likely to shrink.  However, change is likely.
In particular, some of the trends in research that
doomed old-style industrial research labs~\cite{Odlyzko1995} are
now showing up in universities.  A natural development might
lead to the abolition of traditional departments,
for example.
% It is quite conceivable, for example, that the division into traditional departments may
It will be interesting,
therefore, to see the degree to which universities are willing, or are forced, to
outsource their library functions to Google and publishers,.
That might that serve as an indication of how likely more disruptive
moves are to occur in academia.

More generally, the evolution of scholarly publishing 
provides interesting insights into the evolution of
commerce in general.  The availability
of the comprehensive and reliable ARL statistics 
enables us to obtain quantitative measures of some
phenomena that are usually hidden from view.
One of the most important and interesting of these
phenomena is the growth in price discrimination, as
in the University of Michigan paying \$2 million
for something that the University of Montana
obtains for \$440 thousand.  As was easy to predict
a long time ago, differential pricing
% price discrimination
is the driver behind the systematic destruction of privacy.
This is finally beginning to be slowly reflected in
press coverage, cf.~\cite{Clifford2012b,Economist20120630}.
For the most part, though, it is studiously avoided, both in
scholarly venues (such as the call for papers in
the conference~\cite{ECIS2013}) and in most newspaper articles,
which cite arguments that data sharing rules should not
be so tough as to ``hinder or undermine the ability of companies to innovate''
without explaining what kind of innovation is meant~\cite{Singer201302}.
The reason this topic is so studiously avoided
is likely because it is even less pleasant to think about
than downsizing libraries in order to preserve
Elsevier's profits.  The logic of growing price
discrimination leads to visions of a ``dystopian future''~\cite{Kuchera2013},
where many of the basic assumptions about the economy
no longer apply.
% world (that the economy is based on markets, for example) are destroyed.  As a result, 
Scholarly publishing provides useful insights into
why differential pricing is spreading, why it is spreading so slowly
and surreptitiously, and why its spread is hard to resist.

Scholarly publishing also illustrates some other developments
in the modern economy.  The ability of consumers to
find information and hunt for the best deals is justly
celebrated~\cite{Clifford2012a,Economist20120519}, and 
has done much to improve the efficiency
of the economy.  But this development is accompanied by countervailing
tendencies, such as greater opaquenes in other areas of the
economy, as was evidenced by all the ``toxic'' financial instruments
that facilitated the bubble that led to the crash of 2008.
In journal pricing, we also observe greater opacity, with
list prices being essentially meaningless, and real prices
hidden behind non-disclosure clauses in sales contracts.  

Some of the phenomena that scholars object to, such as publishers
making profits out of the unpaid labor of authors, editors, and
referees, are of ancient standing, and are now being replicated
far more widely elsewhere in the economy.
Many of the most successful tech companies of the modern era,
such as Facebook and Google, derive their value from the unpaid
labor of their users.  Thus what is claimed to have been
a possibly passing anomaly in scholarly publishing is actually
becoming a central feature of the modern economy.

The core of this paper is Section~\ref{sec:bigdeals}, which
explores the effect of ``Big Deals'' on access to scholarly
information among ARL members.  First, though, a general overview
of scholarly journal cost structure and of the perils of
technological and market prognostication is presented
in Section~\ref{sec:futurology}.  Then there is discussion
of the value of quantity and quality in Section~\ref{sec:value},
since many scholars dismiss concerns about publishing by
claiming that all that is needed is a tightening of
standards in order to eliminate any crisis.  Then comes
Section~\ref{sec:bigdeals}, as mentioned above.  It is
followed by Section~\ref{sec:arl}, which describes the
data sources that are used, and their limitations.
Section~\ref{sec:universities} has a deeper look into
the budgets and relative roles of universities,
publishers, and libraries.  Section~\ref{sec:lowcost}
summarizes the arguments that scholarly journal publishing
can be carried on at a far lower cost than it is at
present.  Section~\ref{sec:libpub} delves into the
competition between libraries and publishers, and 
what their best courses of action for the future are.
Section~\ref{sec:economyfuture} is the first of several
sections on the development of the entire economy and
how it relates to what we see in scholarly publishing.
There is discussion of growing opaqueness, of the
development of a ``Tom Sawyer'' trends (as in Tom Sawyer
getting boys in his neighborhood to pay him for the
``privilege'' of doing Tom's chores),
and of growing price discrimination.  Finally,
Section~\ref{sec:conclusions} has the conclusions.

\section{Open Access, academic publishing, and academic inertia}\label{sec:futurology}

% \section{Open Access and the perils of futurology}\label{sec:futurology}
% Scholarly publishing overview and fate of prognostications}\label{sec:futurology}

I first became seriously involved in investigating
scholarly publishing two decades ago, and concluded that
some form of Open Access as well as new, lower cost, publishing models
were desirable and inevitable \cite{Odlyzko1994}.  
Although I still feel they are desirable, it now appears far
more questionable whether they are inevitable, at
least in the form envisaged.  The posting of preprints,
while still not universal, is spreading, and the
arguments for it are even stronger than before.  However,
the high cost journals are still surviving, and may continue
to do so.

This paper is not about Open Access.  Rather, it is about
what is actually happening in the rest of scholarly publishing, and
why instead of pure Open Access, we might instead end up
with something that might have a similar effect.  In particular, we might have
various institutions paying subscription fees to
high-cost publishers to provide access to wide classes of readers
(potentially entire nations).
Thus practically nothing will be said about the extensive 
literature on Open Access, on Green versus Gold models, and the like.
Just a few references to that
area are Peter Suber's blog 
and~\cite{BjorkRL2008,Harnad2007,Harnad2011,Houghton2011,HoughtonW2013,Morrison2009,Morrison2012,SolomonB2012,SwanH2012}.  
There is rapid growth in Open Access, as shown by some of these papers,
by the Directory of Open Access Journals
maintained at Lund University, and by other sources.  
The question is, is it fast enough to prevent publishers from
entrenching their high-cost journal model.
Thus, for example, in mathematics, arXiv receives about 25,000 submissions
per year, but that is still only about one quarter of all mathematics
papers that Mathematical Reviews processes each year.  The growth rate
in arXiv submissions is around 15\% per year, a rate that is far
faster than the approximately 3\% per year growth in the number
of articles, and not dissimilar to other growth rates observed in
other aspects of Open Access.
However, that growth is not as fast as what can be seen in Fig.~\ref{fig:serials}
over the last few years for increase in access to serials at ARL libraries.

While there is no space in this paper for a complete review of the
scholarly literature, it will be handy to build the presentation
and analysis around my own earlier analyses and predictions for this subject,
which provide useful background information on some of these issues.
The first paper~\cite{Odlyzko1994} estimated that a typical article
cost about \$20,000 in author's time and \$4,000 in time of reviewers,
both costs that don't show up in any budgets, but do reflect the
effor devoted to scholarly publications by unpaid experts.
It also estimated publisher revenues per article of about \$4,000,
and internal library costs of about twice that.  It also
estimated that \$300--1,000 per paper would suffice for
very good quality in an online-only environment, and that
costs of digitizing old materials would be low (see also~\cite{Odlyzko1997}).
All those estimates still appear reasonable, and even
lower costs have actually been demonstrated, as is
discussed in Section~\ref{sec:lowcost}.
These estimates were then used to argue for a shift to
new low-cost and Open Access journals.

While the conclusions and predictions of~\cite{Odlyzko1994}
did reflect a realization of how much of an obstacle the inertia
of the academic system posed, the main failure of that paper
was in underestimating
that inertia.  An explanation (even if not an adequate
excuse) for this was that the paper was written at an
industrial research lab, so with an inadequate appreciation
of how slowly everything changes in universities.  I forecast
in that paper that the high-cost journal system would 
not change materially for a decade, but would likely collapse
within two decades.  Had I had my current experience as
a professor and university administrator, I would likely have
doubled both estimates.  As has been noted many times,
``scholarly behavior
is profoundly conservative where communication is concerned and that, if
anything, younger members of a discipline are even more conservative than
their elders who are better established''~\cite{Mabe2010}.
The growth in novel forms of peer review has been extremely slow.
Even the high energy theoretical physicists, the first field to
embrace the predecessor of arXiv, and to submit virtually all 
their preprints there, continue to publish most of
their papers in conventional journals.  
A particularly telling observation is that even Stevan Harnad, that indefatigable
advocate for Open Access, has scaled back his effort in this area in
frustration.

Subsequent studies of rates of diffusion of technologies found that while new forms
of communication were often embraced rapidly by scholars~\cite{Odlyzko2002},
that was not the case with the basic journal publications, which were
often deeply embedded into the sociology of their fields, and so on
their own would likely take generations to evolve substantially.
The only way to speed up the process was to have ``forcing agents,''
who can compel change~\cite{Odlyzko1997b}.  In this area I expected university administrators
to take the lead.  However, with a few exceptions, they have been slow to act or lead 
either from ignorance, or pusillanimity, or perhaps from a deeper
understanding of how fast academia can move.  As a result Open Access mandates
have been coming primarily from funding agencies, and relatively slowly.

The high internal costs of libraries compared to the purchase cost
of serials was an obviously important factor~\cite{Odlyzko1994,Odlyzko1999}.
Further, consideration of the entire scholarly communication system,
in which publishers and librarians were the intermediaries between
scholar authors and scholar readers led to the prediction that
there would be an intensifying conflict between these two groups,
as old roles became obsolete or were shifting, and new roles were
emerging.
From the beginning it seemed likely that publishers would be more successful at it than
librarians~\cite{Odlyzko1999}.  The evidence of this paper supports
that prediction.  However, the field is in a state of ferment, and
it is hard to tell how the competition will turn out in the end,
given that Open Access mandates are increasing, and new inexpensive
journals are proliferating.

% tragic loss:  KingMR of 1977 had estimates 

%        authors     12%
%        publishers  14%
%        libraries   10%
%        users       64%

% 20 K for research
% 4 K for refereeing
% 1 K for review
% 4 K for publisher revenues

% \section{Slow progress of Open Access}\label{sec:OA}

\section{Quantity, quality, and value}\label{sec:value}

The basic assumption made in this paper is that more information is better.
We do have a rising chorus of complaints about information overload.
Furthermore, most researchers depend most of the time on a limited
number of information sources.  This is the basis for some of the
concerns about the viability of the ``Big Deals'' (see, for example~\cite{AspesiS20110419}).
Research libraries might reduce their subscriptions to just
a small core of journals, and rely on purchasing access to individual
articles as needed.
Just how much are all those extra journals worth?  
Why can't the libraries be more selective?

There are several answers to that.  There is a continuation of
the traditional growth in the volume of scholarly information,
at about the traditional 3\% per year rate that has prevailed
for many decades.  There is no sign this is going to stop.
Furthermore, with the increased emphasis on interdisciplinary
and multidisciplinary research, there is a need for access
to all the research that is being produced.

Second, all the evidence we have
is that libraries are not very good at selecting the best.
And, of course, the scholarly community itself is not very good
% very poor
at deciding what is the best.  The defects in the editorial and
refereeing processes have been known for long, 
and are becoming ever more apparent. 
Scholars cannot rely on just a narrow slice of available
information even if it is confined to what has received the highest grade of approval
from the current system.
% use the full spectrum of information sources that are available.

Another answer is that all those extra journals must
be worth something.  This is demonstrated by the behavior of
scholars and libraries in the past.  After all, the bottom quarter (in terms of
serials received) of ARL members obtained fewer than 17,000
serials in the 1999--2000 academic year, before ``Big Deals'' became
dominant, while the top quarter all procured over 34,000.  
Why did those richer libraries bother?  If the extra 17,000 serials
could be dispensed with, why didn't those institutions save the
millions of dollars that each one spent annually on
acquiring them?

Further, decisions about usage, acquisition, and retention of scholarly
material are made at the margin.  And most of the current physical
collections are of marginal value.  One can imagine that in 
negotiations, in response to librarians' pleas that 
the ``Big Deals'' are unaffordable, publishers point out
that they would be easily affordable if librarians sent off
much of their hard copy material to distant and inexpensive
storage facilities, shut down some library facilities, curtailed operating
hours, and so on.  All these steps would lower the level of access
to useful information, but so would abandoning the the ``Big Deals.''
The real question is, which hurts more?
% The value of a doubling of the
% number of serials that are available might seem slight, but then
% slight is also the value of the most of the books and old serials
% that are on library shelves.  And those can be sent to some inexpensive
% storage facility offsite, saving money that can be used to pay for
% the new serials.  
In either case, resources are devoted to providing access to material
that, for the most part, has little value individually, but substantial value in total,
not least in the ``option value'' of being able to access it when
desired.

% If interlibrary loans or individual article purchases are to satisfy researcher .  Which is

And of course there is much more that can be said in favor of wider
availability.  The Open Access movement is based on the
argument that we need far wider distribution of research results.
I won't attempt to survey all the arguments that have been produced, except to say
that they are overwhelmingly persuasive, and are becoming ever
more valid as scholarly literature grows and the need for
and opportunities in interdisciplinary work increase.

Bradford's law from information sciences was formulated to provide
a quantitative estimate of diminishing returns from search in journals.
This ''law'' (which is an empirical observation, akin to various
other ''laws,'' such as those of Pareto and Zipf), together with
other evidence from a variety of fields, suggests a more general
rule~\cite{Odlyzko2012} that as a very rough
guide, information, whether in a library, or on a hard disk, or
measured in transmission capacity, should be valued on a logarithmic
scale.  Thus if a collection of 10,000 serials is valued at 4,
one that is 10 times as large, with 100,000 serials, might be worth 5.
That is a gain, but not by factor of 10, but just by 1.25.  
For this paper, though, a precise measure is not necessary.
Wider access is taken as positive.
What we find, as is shown in the next section, is that
``Big Deals'' have led to a major step in the right direction.

\section{Effects of ``Big Deals''}\label{sec:bigdeals}

% Fig.~\ref{fig:serials} shows that the median number of serials
% received by ARL members has more than tripled in the last half
% a dozen years.  Curves for the 1st and 3rd quartiles all show
% dramatic growth, with higher growth for the 1st.  This
% compression is exhibited more explicitly in Fig.~\ref{fig:ratio}.

\begin{figure}
\centerline{\epsfig{figure=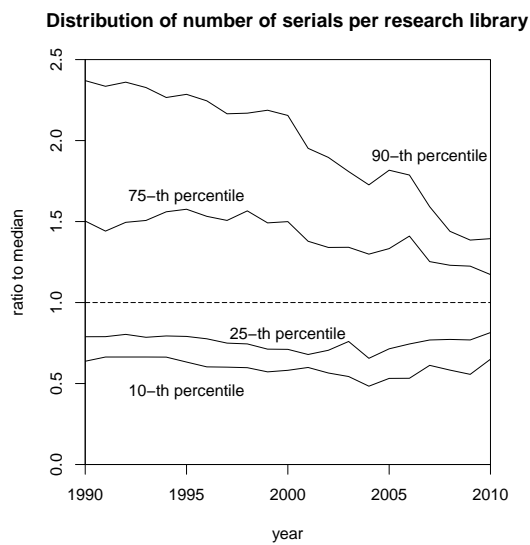,height=10.2cm,angle=-90}}
\caption{Ratio of the number of serials received by various research libraries
to the median, for 10th, 25-th, 75-th, and 90-th percentiles, ranked
by the number received.}
\label{fig:ratio}
\end{figure}

ARL statistics for the academic years 1989-1990
through 2009-2010 (depicted in all the graphs as 1990 to 2010),
show that the last decade
has produced a remarkable increase in availability of serials.
% (again, this is just for persons affiliated with ARL members).
This is shown graphically in Fig.~\ref{fig:serials}, which
is based on a ranking of the ARL academic libraries by the
number of serials they receive.

\begin{figure}
\centerline{\epsfig{figure=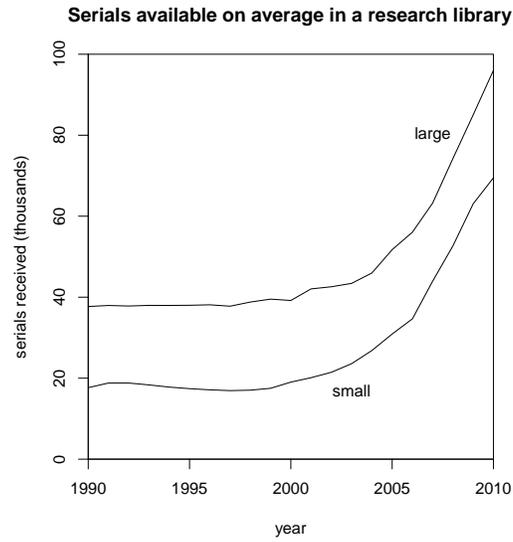,height=10.2cm,angle=-90}}
\caption{Average number of serials received in smaller and 
larger research libraries, 1990-2010.}
\label{fig:aver}
\end{figure}

\begin{figure}
\centerline{\epsfig{figure=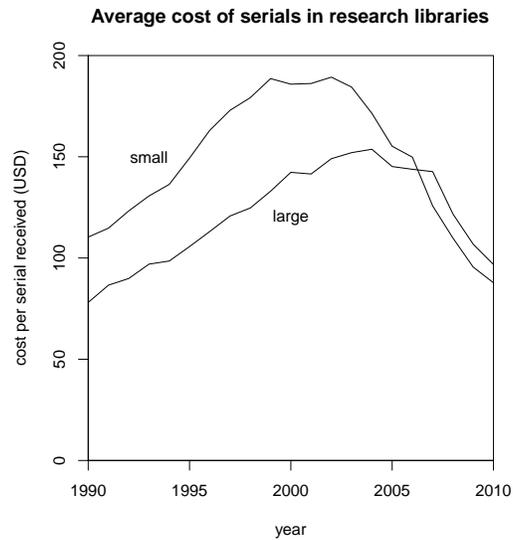,height=10.2cm,angle=-90}}
\caption{Average price per serial received that was paid by the smaller
and larger halves of research libraries, 1990-2010.  (Current dollars, not adjusted
for inflation.)}
\label{fig:price}
\end{figure}

The increase in numbers of available journals was also accompanied
by a notable leveling in availability among ARL members.
This can already be discerned in Fig.~\ref{fig:serials},
but is much clearer in Fig.~\ref{fig:ratio}.  For each
year, the academic members of ARL were ranked by the
number of serials they received, and the percentiles and
displayed ratios were computed.  There is a great degree
of compression, especially at the top, with the 90-th percentile
institutions far closer today to the median than in earlier
years.  The distribution of library budgets has not changed
much over this period\footnote{See Section~\ref{sec:arl}.}, 
so what we observe is the result of
increased price discrimination.  Figures~\ref{fig:aver}
and ~\ref{fig:price} show statistics for the ``small''
vs. ''large'' ARL libraries, defined in each year as those
below and above the media in terms of total library budgets.
(Thus ``small'' is a relative term, as in 2010 this
category included a few libraries with budgets of over \$22 million.)
They demonstrate that the ``small'' institutions gained more
than the ``large'' ones\footnote{However, this observation should
be treated with caution.  Library budgets and purchasing
patterns depend heavily on the nature of the institution,
for example whether it has a medical school, or a law
school, and so on.  Therefore it would not necessarily
be justified to conclude that the higher prices that
``small'' institutions paid per serial in 1990 compared
to ``large'' ones represented price discrimination.
However, the relative gains visible in Fig.~\ref{fig:ratio}
do appear to clearly favor ``small'' schools.}.
The average price per
serial received by ARL academic members has plunged
over the last decade, to a level last seen around 1990,
even in current dollars, without any adjustment for
inflation.
(Unless stated otherwise, all dollar figures in this
paper are in current dollars.)

The growth of the last decade was certainly made possible by
improvements in technology, so that huge collections of
journals can be made available in electronic versions even
by libraries that could not possibly hope to house them
physically.  But much of the credit must be assigned to
the reviled ``Big Deals,'' in which publishers enriched
their offerings by tossing in many additional journals.
This increase in journal availability was not accompanied
by commensurate revenue increases.  As is shown in the ARL
statistics summarized in Section~\ref{sec:arl}, while the
growth in spending on serials did outpace the growth
in library budgets, it did so by a smaller margin in the
first decade of the 21st century than in the last of the 20th.

A serious deficiency of the results presented here is
that they tell us practically nothing about the ``Big Deal'' practices
of the large publishers, such as Elsevier, Springer, and Wiley.
Those are the publishers who attract the most attention,
and they also take a lion's share of the journal spending
at libraries.  However, they publish relatively few journals.
As an example, Elsevier, the largest of these publishers, 
has just under 2,000 serials~\cite{Elsevier2011}.  They
are disproportionately expensive, and also have disproportionately
many articles.  However, since ARL statistics only provide
the total number of serials in member collections, the
contribution of Elsevier is essentially in the noise.
What we can observe is basically just the effect of various aggregators.
Furthermore, many, and perhaps most, scholars, care only about
peer-reviewed journals.  There are only about 20,000 to 25,000
of them, and the ARL statistics this paper is based on do not
distinguish them from others.

Publishers' pricing practices are not the result of altruism.
These practices make excellent business sense for publishers.
With electronic delivery, providing access has
practically zero marginal cost.
From long interactions with libraries, publishers likely have a good
sense of how much they can squeeze out of each institution.
Once that point is reached, the incentive is to increase
usage by everyone at that institution.  That way people
get introduced and hopefully addicted to new data sources
from that publisher, which will make it that much harder
for their libraries to pull out of the ``Big Deals.''  Further,
making all journals available to all customer libraries
serves to decrease the advantages of Open Access.
The faculty, students, and staff have the necessary
articles seamlessly available (courtesy of the librarians
and the library budget), and so are less likely to complain
to their colleagues at other institutions about not being
able to find all they need.  Therefore those colleagues feel
less pressure to get involved in Open Access activities.

In the short (and even intermediate) run, ``Big Deals'' therefore
do promote wider access to scholarly literature, and they do serve
to push libraries to be more efficient.  
Thus what we observe in the
ARL libraries is Adam Smith's ``invisible hand'' producing
socially desirable outcomes.
In the long run, of course, ``Big Deals'' do entrench the
publishers, their profits, and their inefficiency.

Perhaps the greatest puzzle about the wider availability
of journals through ``Big Deals'' is that it has not
occurred much earlier.  This has implications for the rest of
the economy, and will be discussed later.  First, though,
let us look at the ARL data, and at some general financial
statistics of universities, libriaries, and publishers.

\section{ARL and other statistics}\label{sec:arl}

% This section provides details about the main sources of data used for this study.

The Association of Research Libraries (ARL) has, as of this writing
in early 2013, 126 members.  Of these, 11 are public, government, or
other non-profit non-academic institutions (Library of Congress,
Library \& Archives Canada, New York Public Library, ...).  Those 11 are excluded from
all statistics presented in this paper, since they usually have
much wider missions, and are on average far larger in terms of
budgets.  (The largest budget among ARL members has for a long time
belonged to the Library of Congress, which currently spends about
6 times as much as the largest academic library, that of Harvard.)

ARL prepares annual reports on operations of its members, which
are available through~\cite{ARL}.  They cover academic years from
July 1 to June 30, and have in recent years been prepared by Martha
Kyrillidou and her colleagues.  Thus the report entitled 
{\em ARL Statistics 2009--2010} covers the 2009--10 academic year,
and was issued in the fall of 2011.  It will be referred to in
this paper as the ARL 2010 report, and similarly for other
years.  Almost all statistics in this paper are based on the
1990 through 2010 reports.  The 2011 report is now available,
but the bulk of the research was done before it was published,
so it is not included.  However, it does not appear including it
would have led to any significant changes in either the statistics
or the conclusions.

The membership of ARL is very stable.  Over the 1990--2010 stretch,
academic members grew from 107 to 115, with just one withdrawal.
Although the non-renewing library (Stanford) was large, and
the 9 new members considerably smaller, the distribution of
library budgets has not changed much.  In 1990, the average
budget was \$12.99 million, and the standard deviation was 50.0\%
of that.  By 2010, the average budget had grown to
\$27.49 million, while the standard deviation only increased to 54.2\%,
and the distribution curves for budgets (not shown here) were almost identical for those two years.
(All figures are in current US dollars, not corrected for inflation.  Canadian data
is presented in US dollars in ARL statistics.)

ARL statistics are an unparalleled resource of detailed information about
research libraries.  They go back to 1907 (the famous ``Gerould Statistics'' of the
first few decades).
They are very carefully collected, and are a result of the very
open and collaborative culture of librarians.  Still, in spite of strenuous
attempts at uniformity (and use of National Information Standards Organization 
guidelines), they are not perfect.  The voluntary nature of reporting, the presence of
legacy systems and procedures, and sudden local changes leads to 
some obvious anomalies.  As an example, the expenditures
by Harvard on acquisition of serials reached \$10.5 million in the 2004
report, then came down to \$8.4 million in the 2009 report, and then
skyrocketed to \$15.3 million in 2010.  Yet another anomaly was the
number of serials received by the University of Michigan, which was
reported as 67,554 in 2004, 124,809 in 2005, 118,654 in 2006, and 71,788
in 2007.  Both cases likely arise from changes in categorizations.
Such anomalies were ignored in preparing the statistics in
this paper, in the expectation they would not affect overall averages
much.  (In a few cases, missing values were interpolated.)  
Anomalies of this nature do not appear to be present in total library
budgets, nor in the amounts spent on acquisitions.  This is likely
the result of stricter accounting standards and controls for
monetary expenditures.

The serial counts shown in the figures are those for the total number
of serials reported as being received by various libraries.
The serials that are purchased are a more relevant figure.  However,
not all libraries report this figure separately, so in order to have fuller
coverage, the total serial count was used.  That is not likely to
cause a major distortion, as among those institutions that do
break out the purchased serials figure, the fraction of serials
that were received but
not purchased was 24.1\% in 1990 and 27.6\% in 2009\footnote{There were
various other factors that were disregarded, as they did not
appear to lead to large changes in the relevant statistics.
For example, starting with the 2007 report, ARL members were
asked to report serial titles that were received, and not
subscriptions.}.

For all statistics, calculations were done separately for each
year.  Thus the ``small'' category of libraries in Fig.~\ref{fig:aver}
included all libraries whose total budget was at or below the median
for all budgets for a given year.  

One of the major deficiencies of ARL reports is that they seriously
understate the internal costs of libraries.  They count only the
direct pay of employees, and not any of the health care or retirement benefits those
employees receive.  ARL reports also do not include most of the
maintenance costs of library facilities (heat, electricity, cleaning, ...).
Given the enormous variation in how universities 
account for such costs, it would indeed be hard to come
up with a uniform reporting standard that could be implemented
easily, so these omissions are understandable.  Still, what
they mean is actual total library costs are likely about 25\%
higher than the budgets given in the ARL reports (as confirmed
by some correspondence with librarians).

\section{Universities, publishers, and libraries}\label{sec:universities}

The ARL has statistics showing library budgets as fractions of total
university budgets for a sizable collection of their members~\cite{ARL2}.  
The chart for the 40 members that have reported since 1982 shows an
inexorable decline in this ratio, from about 3.7\% to a bit under 2.0\%\footnote{It
is worth pointing out, though, that the chart for the 17 members that
reported since 1966 shows an initial jump from a level of about 2.9\%,
and then essentially a constant fraction until about 1980, when the
decline starts.}.  As was mentioned above, more accurate accounting
would inflate the cost of libraries by about 25\%, so the decline
was from a level of about 4.5\% to about 2.5\%.  But the trend
and size of the decline would not have been affected materially by this
adjustment.

Thus libraries have been getting a shrinking part of the entire
budget, and there would be no journal crisis had they just maintained
their share from the 1970s.  However, this has to be put in context.

Universities have an insatiable desire for more funding.  They are
complicated entities, and need to support many activities.
Retirement bonuses for administrators, buyout packages for heads of athletic
departments, fancy big football fields, and of course
fancy salaries for football coaches, all compete for funding
with a variety of other categories, including libraries, faculty salaries, 
classroom and lab buildings, mental health counseling for students, and many other
cost centers.
% \footnote{Large libraries pay on the order of \$2 million
% per year for access to all Elsevier journals.  That would not pay for
% a football coach of more than mediocre caliber.}.
Note that the \$2 million or so that a typical large university
paid for access to all Elsevier journals in 2007 would
not be enough to pay for a single prominent football coach.

A very remarkable fact is that, in spite of all the wailing and
gnashing of teech,
universities have been extremely successful in
obtaining more funding.  
This is surely the result of the observation cited earlier
in the Introduction, that education is becoming ever more important
as technology and the economy advance.
Table~\ref{table:growth} shows 
statistics for ARL members in the last two decades (with CPI denoting
the main Consumer Price Index for the U.S.)\footnote{The GDP growth rates are
derived from \\
$\langle$http://www.usgovernmentspending.com/us\_gdp\_history\#usgs101$\rangle$,
the CPI ones from \\
$\langle$ftp://ftp.bls.gov/pub/special.requests/cpi/cpiai.txt$\rangle$.
The other figures are from ARL compilations.  Although ARL has many Canadian
members, this presentation is U.S.-centric and ignores the fact that Canadian data
is part of the ARL reports.}.
Their budgets have grown far faster than the economy as a whole.
And so have library budgets, even though they grew more slowly,
and not as fast as spending on library acquisitions.

\begin{table}[htb]
\begin{center}
\caption{Annual compound growth rates.}\label{table:growth}
~ \\
\begin{tabular}{llrrrrrrrrrrr}
year  & ~~ & US GDP & ~~ & US CPI  & ~~ &  university  & ~~ & library  & ~~ & library  & ~~ & serials \\
      & ~~ &        & ~~ &         & ~~ &  budgets     & ~~ & budgets  & ~~ & purchases & ~~ & purchases   \\ \hline
1990--2000    & &     5.55\%  & &     2.85\%  & & 6.83\%      & &       4.65\%    & &    6.16\%   & &    6.93\%     \\
2000--2010    & &     3.85\%   & &    2.53\%  & & 8.70\%   & &    3.74\%    & &   4.84\%   & &    5.64\% 
\end{tabular}
\end{center}
\end{table}

[Note: The growth rates for total budgets for ARL member universities are suspiciously large,
and will need to be verified.  They are based on data from the ARL for 1990, 2000,
and 2009 (since figures for 2010 were not available, so a 9-year annual growth rate
was entered in the 2000--2010 line).  Only figures for the 89 ARL members for
which data were given in the ARL tables for each of those three years were used.  
Some of the figures in those tables seem anomalous.  For example, the budgets of the University of
Connecticut, University of Michigan, University of New Mexico, Ohio State, Temple,
University of Utah, University of Virginia, and Western Ontario are all shown
in those ARL compilations
as increasing by more than a factor of 3 from 2000 to 2009.  This is rather implausible,
unless it corresponds to some major changes in accounting, such as sudden inclusion of
revenues from affiliated hospitals, and if it does, this would be misleading.  
Still, there have been other reported statistics of higher
education budgets in the U.S. outpacing not just inflation, but GDP growth and
sometimes even medical sector growth, so whatever the right figure in that
column, it is likely to be high.]

\begin{figure}
\centerline{\epsfig{figure=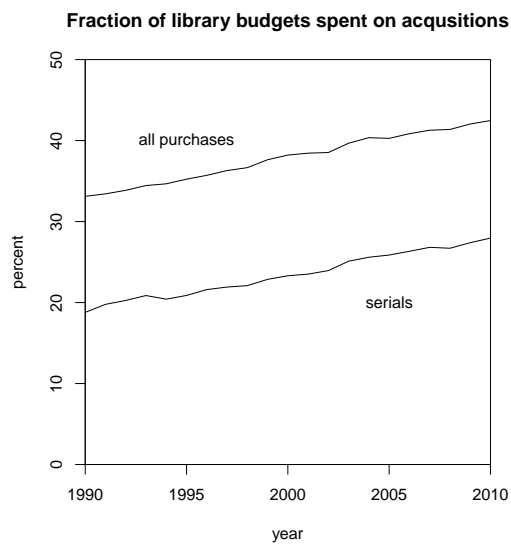,height=10.2cm,angle=-90}}
\caption{Fraction of library budgets devoted to all acquisitions
and to purchases of serials.}
\label{fig:buy2}
\end{figure}

One result of the trends shown in Table~\ref{table:growth} is displayed
in Fig.~\ref{fig:buy2}.  The share of library budgets that goes out in
purchases of books, journals, and databases has grown substantially,
from 33\% in 1990 to 42.5\% in 2010.  (If we correct for the undercounting
of library budgets cited before, the growth is from about 25\% to about 33\%.)
Further, all of this growth is accounted for by serials.  Books and other
materials have just about held their own (with books shrinking at the expense
of the rest).

Collectively, academic members of ARL had 
budgets that added up to \$3.16 billion according to the 2010 report,
so that their actual costs
were likely close to \$4 billion.
Of this, \$1.34 billion was spent
on outside acquisitions, and \$884 million of that on serials.  

The entire scholarly publishing market, as measured in terms of publisher
revenues, is someplace around \$20 billion per year\footnote{See \cite{WareM2009}
or the Jan.~6, 2012, announcement of the Simba Information report, \\
$\langle$http://www.simbainformation.com/about/release.asp?id=2503$\rangle$.
Many sources look only at STM publications, but one needs to consider
also the humanities, etc., which, while not large in terms of revenues,
are not negligible.}.
Of this, journals account for \$8-10 billion.

There are various estimates for the number of scholarly articles  
published each year, although it does seem to be agreed this number
grows by between 3\% and 3.5\% per year.  The report~\cite{WareM2009} had a figure of
about 1.5 million English-language peer-reviewed being published in 2009.
There are some estimates that are higher, around 2 or 2.5 million, and
they may also be correct, in that they may include non-English publications
as well as some serials that are not peer-reviewed.  
This paper uses the round figure of 2 million for simplicity.

% I will use the figure of 2 million, especially since 
% many publications that are not peer-reviewed are still useful.

% publishers: can move to lower cost models and keep profits

% publishers: likely to support deals on orphan works, and other
%   measures that would reduce justification for libraries to
%    maintain their expensive infrastructures

% national licensing, analogies with pharmaceuticals

\section{Lower cost models}\label{sec:lowcost}

Revenues of \$10 billion and 2 million articles mean an average
cost to society of the scholarly publishing system of about \$5,000
per article.  This does seem to be typical.  
Note that this does not differ much from the estimate
made almost two decades ago~\cite{Odlyzko1994} of about \$4,000 per article.

If we consider the
Reed Elsevier annual report for 2011~\cite{Elsevier2011}, we find
that the science and technology segment of the Elsevier business
unit had revenues of about \$1.7 billion, and published around
240,000 articles.  Since that division also published books
and engaged in other activities, if we assume \$1.2 billion
as the journal revenues, we obtain a figure of about \$5,000
per article.  
Many professional societies operate at similar levels.
For example, the American Mathematical Society had journal
revenues of \$4.7 million in 2010, and published 873 articles,
for revenues of \$5,400 per article\footnote{Figures derived from
the 2010--2011 Annual Report~\cite{AMS}.  (A more accurate
accounting would almost surely produce a somewhat lower figure,
though, as the revenue figure includes services provided to
other scholarly organizations.)
That report shows total publications program revenues of \$19.5 million
and expenses of \$14.7 million, for a profit margin of 25\%.
This appears to be not atypical of nonprofit professional societies,
which often use profits from publications to subsidize their
other activities.}.

Most of the arguments for Open Access are valid irrespective of the costs of publications,
and are based on the public good, efficiency of research, and similar considerations.
However, the possibility of moving to dramatically lower cost
structures does make a switch to new business models much easier to perform.
It has been clear for two decades that much lower costs in
scholarly publishing are possible, but with some exceptions, 
little has been done to the bulk of the literature to move in
that direction.

As a point of contrast and information, the costs of running the
arXiv preprint server are under \$10 per article that is submitted
each year.  There is still
some cost.  Various problems arise with submitters (most handled
by unpaid volunteers who do some minimal screening to eliminate
inappropriate postings), basic software changes have to be handled, new features
need to be put it, compatibility has to be preserved, and so on.
However, with minimal human intervention costs can be
very low.  Once experts get involved, costs can mount up quickly.
As an example, the MathSciNet reviewing operation of the American
Mathematical Society~\cite{AMS} collects revenues of about \$100 per
reviewed paper, and, depending on how various expenses are allocated, costs
possibly not much more than half of that, or about \$50 per paper.
This time subject experts with PhDs in the appropriate disciplines
are involved, as well as myriads of unpaid reviewers, and publishers
cooperate in providing metadata and other information.    
Still, this cost is about two orders of magnitude lower 
than current average cost of an article.  That serves to disprove
frequent claims of publishers that metadata provision is expensive.

Can one operate high-quality scholarly journals for less than 
\$5,000 per article?  That this issue is still debated is a good
introduction to a major theme of the rest of the paper, the
slow rate not just of actual change, but even of information diffusion in our society.
Almost two decades ago various investigators estimated that 
between \$300 and \$1,000
per article should be sufficient for good service~\cite{Odlyzko1994}.  Since that
time mountains of evidence have accumulated to support that
conclusion.  There are now numerous journals operated
by their editors, without any explicit monetary flows.
Further, we have, among others, the Edgar and Willinsky study~\cite{EdgarW}
of almost 1,000 journals that use their Open Journal System
(in use by about 5,000 journals in all) which found the first
copy costs (which is essentially all that matters if one considers
purely electronic publishing) of \$188 per article.
Clearly, many of these low-cost ventures do benefit from
implicit subsidies (such as free use of university information technology
infrastructure), and would not scale to larger journals without
having to pay for various types of administrative assistance.
Still, when we have examples of 10x cost savings, even a doubling
of the cost to pay for all those extras still leaves us with a 5x cost advantage.

That lower costs were possible was obvious even three decades
ago, since the costs per article varied wildly between publishers.
This showed that costs were not a matter of unavoidable necessity,
but of market power, choice, and inertia.  This has become far clearer since
then.  There are many cost reductions that are feasible and desirable.
% and in the end (aside from some exceptional cases) inevitable.
Let us consider some.

The first step is to abandon print entirely, and publish strictly
electronically.  That will eliminate the expenses not just of the
print operation, but also of the preparation for print.  Once that
is done, it will be much clearer that
there is much more that we can choose to economize on.
For example, why devote a lot of effort to ensuring consistency
in layout, reference styles, etc. in a purely electronic journal?  (Those
activities do consume extensive effort even in low-cost
serials, see, for example, the data for the Open Journal System
journals~\cite{EdgarW}.)
In a printed
version, it might have made some sense, but in the modern online
environment, where readers follow threads from one journal to
another, this uniformity contributes little.  

Some of the terrible waste that is involved in scholarly publishing
involves no money, but extra work for authors.  For example,
some journals insist that not only the final accepted versions of
papers adhere to their chosen format, but that even initial
submissions do.  This is not just wasteful, but silly.
Yet it persists, and provides yet another illustration of the
inertia in academic publishing.

To get to the lowest ranges of costs,
one can also eliminate most of copy editing.  The quality of
author-prepared manuscripts that I find at the arXiv and SSRN
preprint repositories is almost always completely adequate from my personal
perspective.  (The same is true of the conference proceedings
that ACM and IEEE publish, where the authors are responsible
for producing the files using the publisher-supplied style files.)
As a reader of scholarly journals (and I read a lot, in a variety
of fields) I find copy editing useless, although admittedly this is an
extreme opinion.  

As an author,  I find that copy editing subtracts value, by forcing
me to do extra work, usually for no good reason, and often
to correct what the copy editors have done.
% provides negative value.  
% I can tell (true) stories of copy
% editors who did wonderful jobs, catching nontrivial mistakes.
% But that was decades ago.  Today, the quality of work one
% finds is different.
As just one example, a recent set of proofs I received
resulted from the publisher editing the manuscript, and,
among other things, changing the reference style from the
one my coauthor and I used in the preprint to the official
one of the publication.  Although I loathe that style,
the move was understandable, as our contribution is to be
a chapter in a book, where uniformity is expected.  However, the editing introduced all
sorts of grammatical and stylistic errors.  The result was
that my coauthor and I spent more time correcting the proofs
than we would have spent getting things right in the first
place, had we been told what was wanted.

Thus my firm conviction is that one can publish scholarly articles
at much lower cost than is done today.  One can do it even if one
does a moderate amount of copy editing, as the Edgar and Willinsky data~\cite{EdgarW}
shows.  

The \$800 million
that just the ARL members alone spend on journals is about \$400 for
each of the 2 million or so articles that appear each year.
And \$400 should be plenty, as was forecast two decades ago,
and as is demonstrated by many respectable journals.  Yet the
world is currently paying about \$5,000 per article.  The question
is whether we can get there from here.

\section{Competition between libraries and publishers}\label{sec:libpub}

The evolution of academic information systems is viewed most fruitfully
as dominated by competition between publishers and libraries.  
However, one should not neglect the role of an increasing variety
of other players, as this sector evolves.  A century ago, libraries
and publishers were basically the only significant players,
and there was a strict separation in their functions.  Today,
the lines are blurry.  It used to be that publishers put out
books and journals, and libraries preserved them for posterity.
Publishers might keep some copies of old issues in their inventory,
but were not obliged to, and had little incentive to reprint.
But with electronic publications, it is far more efficient
and effective to have the publishers maintain the files, migrate
material to new formats, etc.  (I am ignoring here issues such
as censorship, which is far easier to carry out on a publisher's
few mirrors than it is on a thousand libraries.)  It does not
have to be a publisher that does this, as the Internet Archive
and Google Books demonstrate.  And there are many other players
in this expanding arena,
from high performance computing centers that increasingly 
handle large data sets, to commercial organizations curating
genomic databases.  That is not a surprise.
Information and education are becoming
more important to society, so it is to be expected that more
will be spent on it.  This is the same phenomenon that is
driving up spending on medical care.  Although various reforms
have been carried out and others are planned, the move towards
more of the economy being devoted to health is worldwide,
associated with rising standards of living.  Any increases
in efficiency there are likely to be swallowed up in
new treatments, or just in more treatments for an aging
population.  Similar trends appear likely to operate
in education.  
While the traditional functions of librarians and publishers are obsolete, 
new opportunities are arising for
intermediaries between researchers and other researchers, or
between researchers and the wider public.

% Yet that point of view is strangely absent.  

% The arrival of the Internet is making traditional functions of both libraries
% and publishers obsolete.  This did not mean that there is not going
% to be a role for them in the evolving scholarly publishing industry.
% Their roles have been evolving over time in any case, and the profession
% of librarian only dates back to the late 19th century.  But now change
% has accelerated, and many traditional functions are bound to decline
% or disappear completely.  Who fills the many new information intermediary
% roles is up in the air.  So far libraries have been growing faster
% than the economy as a whole, but not as fast as their universities
% or as publishers.  They have been moving into providing new services,
% but so far their success has been limited.
% (
% and leads to a competition between libraries and publishers~\cite{Odlyzko1999}.  So far the
% publishers have been winning.

\begin{figure}
\centerline{\epsfig{figure=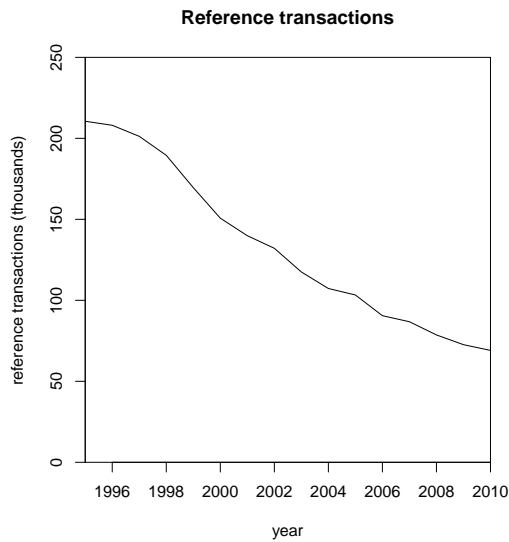,height=10.2cm,angle=-90}}
\caption{Average number of reference transaction in reporting
research libraries, 1995--2010.}
\label{fig:reftran}
\end{figure}

There are many interesting
statistics at~\cite{ARL} demonstrating decline of the traditional
functions of libraries.  Thus between 1995 and 2010, the number
of students at ARL institutions grew by 33\% (with the ranks of
teaching faculty and graduate students climbing 15\% and 43\%, respectively).
The only category of library services involving physical material
that showed growth was interlibrary loans, which climbed 92\%.
This reflects libraries concentrating their budges on serials, and
giving up on trying to keep up with the growth in the number
of new books being published.
In other categories, initial circulation (i.e., excluding renewals)
of physical volumes dropped by 42\%.
Thus it is a gross exaggeration that ``nobody uses the library anymore,''
as one sometimes heard from faculty or students.  But the decline in
borrowings per student by more than half is telling.
What is perhaps most surprising is that the number of requests
for reference assistance dropped by 66\% in absolute terms, as is shown in Fig.~\ref{fig:reftran},
and thus by about 75\% on a per-student basis.
This is certainly a core competency
of librarians, and they are great at navigating the torrents
of electronic information, as well as providing guidance
to the use of traditional printed sources.  However, it
appears that Google, Wikipedia, publisher databases, and
the like are ``good enough'' for most scholars, and that
the convenience of around the clock access from anyplace
outweighs the higher quality that librarians provide.

The declining reliance on reference librarians is intriguing.
In many, perhaps most, of the areas of the economy we see great
willingness to pay more for what is perceived as the very best,
even when the margin over second best is extremely slight or illusory.
So the very top athletes, artists, and lawyers (and sometimes even academics) see their pay
skyrocket, while the crowds of competitors who are almost as good languish
in obscurity and often penury.  Ranks of applicants to top-ranked universities
grow tremendously, even though there is little evidence this leads
to more successful careers.  But librarian skills are apparently not perceived
in this same light.

% Yet not only can the costs of publishing be shrunk, so can
% the costs of current library services.  Those costs are two
% or three times higher than those of purchasing serials.
% The analysis of the library/publisher dynamics in~\cite{Odlyzko1999}
% let to the prediction that publishers would come out ahead, for
% a variety of structural and cultural factors.
% The statistics of Section~\ref{sec:universities} show that
% up till now that is what has been happening.

What librarians could have done, and might yet do, is discussed
later.  First, though, let's consider publishers.  They have moved
faster than librarians to adjust to the new world, as was to
be expected~\cite{Odlyzko1999}.  They have many
options for maintaining their leaderships (as well as their profits)
in the future.  A transition to Gold Open Access (in which authors,
or their funding agencies, pay the publishers and the articles
are make freely available) might not be a threat by itself, but rather
a possible opportunity to increase revenues if one can sustain
subscriptions, say in a hybrid model.
Even if publishers are forced to lower their revenues,
they could maintain their profits by making some of the moves
discussed in Section~\ref{sec:lowcost}, such as giving up entirely
on print, decreasing the amount of copy editing, and abandoning the
requirements for uniform style in each journal.  They might also
increase their profits while keeping revenues steady, by implementing
some of these moves.  

The basic and very promising approach open to publishers is to
continue marginalizing libraries by extending the reach and scope of ``Big Deals.''
The consortium model, in which groups of libraries cooperate
to get access to a ``Big Deal'' is already common, and can
be pushed further.  The ultimate situation might be national
``Big Deals,'' where some top-level bodies pay for access
for everyone from a nation.  Enlarging the ``Big Deal,'' especially
through further mergers, but also by including additional
information sources, can serve to create packages that
simply could not be dispensed with.  The most obvious move
in that direction (which is already taking place to a small extent) is to make books, both current
and old ones, a part of the ``Big Deal.''  (Recall that the
process of digitizing old printed materials is extremely inexpensive.)

While it is a little puzzling that journal ``Big Deals''  
have been so slow to spread, the slow rate for inclusion of books
is much more understandable.
% book situation is different.
For several decades now,
the principal method of using scholarly journals has been
to scan a multitude of articles, photocopy the ones that
appeared to deserve more careful study, and work with those
photocopies.  Hence the transition to all-electronic journals
is relatively seamless, with the photocopying being replaced
by printing.  It is also far more efficient, as one does
not have to go to the library, and handle awkward bound
volumes, and instead prints in one's own office or study.
Even so, there are die-hard traditionalists who continue
to insist that nothing will replace the traditional methods.
However, their ranks are decreasing.  Printed books, on the other
hand, have many more adherents.  They are traditionally
used in ways substantially different from that of articles.
It would have been difficult a decade ago, for example,
to persuade scholars to give them up.  However, improvements
in technology have led to very nice ebook readers and
more general tablet computers, and there is a rapid
move in the population as a whole towards ebooks.
Scholarly publishers therefore have a chance to jump
on the bandwagon and also to speed it up.  By digitizing
their current offerings as well as their old out-of-print
volumes, they can provide better service to readers,
and reduce the need for libraries.
Furthermore, with the improving quality of print-on-demand
services, it is still possible to cater to those who
insist on paper copies.

Making books as well as back issues of journals widely
available in electronic forms could also help in indirect
ways.  In the traditional print world, it would have been
a major disaster for a small institution, say Reed College,
to be given a duplicate of all of the Harvard libraries'
physical collections.  There simply would not have been
the space or personnel to handle such riches.  But, based on experience,
% electronic access is a different story, and, based on experience,
we can be sure that Reed College faculty and students
would use all that is at Harvard if that were available
online.  And once Harvard faculty see their colleagues
at Reed relying on ebooks instead of hard copy volumes,
their resistance to doing the same will diminish.

It would also be wise for publishers to overcome their
reluctance to tamper with copyrights, and to forcefully
push for legislative solutions to the orphan works problem.
Orphan works, which are those that are under copyright,
but whose copyright owners cannot be located, provide
a major justification for the existence of large physical
collections in libraries,
as that is the only legal way to make such works
accessible to the public.  To the extent orphan works
are digitized and made freely available, the need
for library facilities (both physical and personnel)
declines.  That would make it easier for publishers to
appropriate even more of the resources now going to libraries.

What about libraries?  They are handicapped in the competition
with publishers by several factors, see~\cite{Odlyzko1999}.
One of them, that they have the bulk of the resources,
and are thus a fat target, is a strength as well.  
At least in principle it makes possible revolutionary changes.
In particular, as was shown earlier, just the external
journal purchases of the ARL libraries alone could provide
Open Access publishing for the world's entire scholarly
literature.
Had libraries thrown their resources enthusiastically
behind new, low-cost Open Access journals, perhaps the
current scene and the unfolding future sketched here
would have been different.  But that would have required
many research partners willing to put their energy 
into the enterprise (certainly a very doubtful proposition,
given the inertia in the academic system), and the
willingness of librarians to cannibalize 
their bread-and-butter operations.  Certainly
librarians present a classic case of Christensen's
``innovator's dilemma,'' pressed to maintain traditional
services, and therefore slow to embrace new ones.
As an example, digital libraries have been discussed
in the library literature for decades.  Further,
the amount that ARL libraries spend in a single year on
acquisition of serials would have sufficed, with
plenty left over, to digitize all their standard
books and journals that are out of copyright\footnote{As was clear
two decades ago~\cite{Odlyzko1994}, and as has now been
confirmed by various large-scale projects, such as that of the
Internet Archive,
one can digitize a book for about \$20 per copy on average.
Hence the \$884 million spent on serials around 2010, would
have paid for conversion of over 40 million volumes, far more
than the number of unique sets that are out of copyright.
(Harvard reports about 17 million volumes total, the Library
of Congress, 35 million, and this is for everything, including
recent publications.)}.
Yet it was outside efforts, in particular the Gutenberg Project
(the early pioneer, almost forgotten), Google Books, and the
Internet Archive, that led the way\footnote{As another example
of the cautious and incremental librarian approaches, consider
the study~\cite{SchonfeldKOF2004} of costs of print and electronic
journal subscriptions.  It was far too conservative in the
estimates of moving to ``Big Deals,'' in which per-journal costs
outside the purchase price are negligible.}.  
And thus today we do
see libraries moving towards support of Open Access, beginning
to pay some author submission fees, and providing
financial and management support for inexpensive new journals.
But the moves are slow (reflecting, perhaps, the hesitant
embrace of Open Access one detects in~\cite{Harris2012}),
and may not be enough to prevent further publisher encroachments.

We also see libraries moving into other services, such
as providing long-term storage for publications, data
sets, and so on.
However, there they are competing not just with publishers,
who also see the opportunities, but also other organizations,
such as campus information technology units, high performance
computer centers, and a variety of new commercial startups.
The opportunities are many, but so are the competitors.

\section{The future of the economy}\label{sec:economyfuture}

% , opaque pricing and price discrimination}\label{sec:economyfuture}

Scholarly communication and the ``Big Deals'' are interesting
not only in their own right, but also because they provide
interesting perspectives on the development of the general economy.
As one example, one of the reigning mantras of the Internet bubble was 
that of ``friction-free capitalism,'' a term that was
coined by Bill Gates, and one that he was very proud of~\cite{GatesMR1995,Gates1999}.
Barriers would fall, middlemen would be disintermediated (``[o]ften the only
humans involved in a transaction will be the actual buyer and seller,''~\cite{GatesMR1995}, p.~158),
and the economy would be far more efficient.  Yet there was a certain implausibility
from the start about the vision of ``friction-free capitalism.''  How could
it be reconciled with the rich profits Microsoft was making (and continues to
make today, even in the face of competition from free Linux and other operating systems,
and even as PC sales sag)?

There were many arguments that while the economy would get more
efficient, it would depend crucially on ``speed bumps'' \cite{Odlyzko1996}.
% be characterized by ``speed bumps'' (\cite{Odlyzko1996},
(Such speed bumps were actually mentioned briefly by Bill Gates in~\cite{GatesMR1995}, p.~234.)
Those speed bumps might take a variety of forms, from various methods
for increasing switching costs to patent and regulatory barriers.
(It is noteworthy to observe the growth in the ranks of lobbyists in
Washington, especially those representing the technology industry.
When it is only the inertia of habit that induces customers to 
stick to a particular web site for browsing or shopping, any
speed bumps matter, and court injunctions, say, become giant speed bumps.)
Mergers and acquisitions, such as those witnessed in academic
publishing over the last couple of decades, can be viewed as 
attempts to create bigger speed bumps, in which 
% an institution
refusing to deal with a particular publisher becomes unthinkable.
(That lets publishers get away from the constraints imposed
by models such as those studied in~\cite{McCabeSF2012}.)

Among speed bumps, many can be categorized as part of confusology.
Scott Adams back in 1998 had the prescient idea, in one of his Dilbert books, that
``Companies form confusopolies to make it impossible
for the average individual to determine who has the lowest price,'' \cite{Adams1998}, p.~161.
Let us use the word confusology for the more general art of befuddling customers.
It is an art that flourishes, and financial analysts have claimed recently, for example,
that ``[f]or years, the telecom industry has thrived on obfuscation''~\cite{MoffettDC20120718}.
(In view of the high demand for
expert practitioners in confusology, it is strange that academia has
not yet developed courses in it.)
A very elementary form of confusology is to pollute the consumer website data,
where an ``investigation has uncovered fake reviewing on an almost industrial scale,
with companies paying offshore contractors to post numerous glowing accounts of their
activities, yet maintaining they are from unbiased consumers''~\cite{Smith201301}.

%   growth in lobbying, patent lawsuits, ..., symptoms of increased importance of speed bumps

%	TiVo   example of failed attempt to get a chokehold

%	Bruce Schneier:  feudal overlords

With the benefits of hindsight, we can see that the economy has become
more efficient, and many intermediaries have been disintermediated.
There are many news stories about modern technology empowering customers,
cf.~\cite{Clifford2012a,Economist20120519}.
Brick and mortar stores that have not mastered confusology sufficiently
do suffer from shoppers who look and test, and then comparison shop at
Amazon and other places on their smart phones, often right in the store.
But they are likely to soon learn, or else will go out of business.
Two stories in the same recent issue of the {\em New York Times}
illustrate both the problem (for sellers) and the obvious
potential solutions~\cite{Clifford201301,PogrebinF201301}.
One article discusses apps available to customers that alert them to sales
on goods and services they are interested in, to compensate for the
rapidly changing prices~\cite{Clifford201301}.
The other, though, shows how even in the world of high-priced art,
where auctions, which are acclaimed by economists as an ideal tool
for price discovery, are a basic sales method, sellers manage to
confuse the scene, even though the buyers are usually very
wealthy and thus presumably sophisticated individuals~\cite{PogrebinF201301}.
Thus we can expect (and already see in many situations) sellers 
avoiding the disadvantages that the stores in~\cite{Clifford201301}
suffer by moving to offering personalized discounts, etc.

Thus on balance it is not clear that the economy has become more transparent,
or that the role of intermediaries has diminished (even though the dominant
ones may now have different names, like Google, eBay, and Amazon).
Auctions for ads by Google are a triumph for theoretical economics.
But they are imposed by Google, which has an interest in having a
transparent market that will maximize its revenues.
Elsewhere, we find dramatically different pictures.  By far the
largest industry is health care.  It is also among the fastest
growing, and there is no denying that it is improving lives
(although not as fast as it is swallowing resources, it seems).
% at what seems a much slower pace than that of costs).
It is also among the most opaque.  (The discussion here, as
well as in most of this paper, is very U.S.-centric, and
various parts of the argument and evidence may not apply
to other countries.)  An individual, even one not dealing 
with an emergency, who attempts to deal with
this system encounters a bewildering morass, with prices
unobtainable or incomplete, and varying wildly between
different providers, and cryptic codes in bills that conceal
obscure and unexpected charges, e.g.~\cite{Rabin201302}.  Not paying such bills promptly,
though, often leads to encounters with abusive
bill collectors.  A standard way to mitigate such problems,
for those affluent enough, or with jobs that provide
good employee benefits, is insurance.
But that introduces a new (and expensive for society)
level of intermediaries that also are good at confusology.
There are all those ``reasonable and customary fees,'' that
do not necessarily seem reasonable, and whose determination
is a closely guarded secret of the insurers (although a few
lawsuits have led to some reform and transparency).
An experienced business journalist made ``more than 70 phone calls to 16 organizations''
and yet was still not able to find out how much he would have to
pay for a prescription drug that keeps his cancer under control~\cite{Lalli2012}.
Digital health records should eventually lead to better medical care.
However, their spread was pushed by the government largely on the
promise of reducing costs.  As of now, ``[w]e've not achieved the productivity and quality benefits that are
unquestionably there for the taking,'' and
``evidence of significant savings is scant, and there is increasing
concern that electronic records have actually added to costs by making
it easier to bill more for some services''~\cite{AbelsonC2013}.

Medical care is just one example of extensive confusology.
The financial industry is another.  During the Internet bubble,
banks were often seen as obsolete intermediaries that were
at best doomed to shrink.
Instead, along with the rest of the financial
industry, they have prospered.  One of their routes to prosperity was
creation of confusing products.  Acclaimed by economists and
regulators as instruments for spreading the risk to those able and willing
to bear it, they ended up in the hands of those who understood them
the least.  
Even the most visible financial markets, with all the rules and regulations, were visibly
murky even before the crisis of 2008~\cite{PulliamSS2007}, as much
of the trading was carried out in opaque venues, out of sight
of the public.
More recently we have learned that the Libor rate, perhaps the
most prominent financial benchmark, had been manipulated for
years.  The growing complexity of our interconnected world
appears to allow for increased manipulation and inefficiency.
Even regulators, with access to far more information than the general public,
much less the general public, often fail to understand
what is happening.  Large banks, which created and helped conceal many
of the dangerous financial practices that brought the economy to the
brink of precipice in 2008, had to be rescued because they
were Too Big to Fail.  In the process, the surviving ones
have become Much Too Big to Fail,
% (there has been a big jump in concentration of assets in the banking industry in the
% hands of the largest ones), 
and in spite of pledges of reform,
appear to be just as opaque as before~\cite{PartnoyE2013}.

It is not just profit-making corporations that rely on such methods.
Higher education institutions are expert practitioners of confusology.
University budgets are opaque, and little comparative performance
data is available for different institutions.
While affirmative action on grounds of gender
or race does attract substantial scrutiny, the extent of
athletic or ``legacy'' preferences usually only catches public
attention when some scandal erupts.  And higher education
pricing is also increasingly confusing.  While tuition
is rising rapidly, so is ``financial aid,'' which at
private universities in the U.S. now amounts to over 40\%
of tuition revenue.  Furthermore, there is an increasing
amount of bargaining between parents and institutions.

The commercial higher education sector is an extreme
example.  An investigation by a U.S. Senate committee
found that among 30 for-profit colleges, ``an average of 22.4 percent 
of revenue went to marketing and recruiting, 19.4 percent to profits 
and 17.7 percent to instruction''~\cite{Lewin2012}.
Might seem shocking, but figures for non-profits might also
turn out shocking, if they were available.  In general, in many sectors of the economy
the basic cost of providing goods and services is often
dominated by various overhead expenses.

From this point of view, academic publishing is only
catching up to the general trends.  Whereas in the
old print-only world, pricing was fairly simple, now it is
concealed by the non-disclosure clauses of the ``Big Deal'' 
contracts.  And a new group of intermediaries has arisen who,
for a fee, and based on their private knowledge of the market,
help libraries negotiate improved ``Big Deals.''  
Publishers use their superior knowledge of the market as
well as of usage patterns to argue that the deal they
are offering a particular institution is the very best
deal in the world.  (And it often is, as the availability of
many measures makes it possible to custom tailor metrics that
show just about anything one can wish.)
Thus it is hard to tell what is happening.

% While the trend towards opaque pricing is general, and one
% can argue that it has been slow to arrive in scholarly publishing,
% this area has been a pioneer in the move towards that might
% be called ``Tom Sawyer economics.''  In Mark Twain's classic,
% Tom Sayer, condemned by Aunt Polly to a day of hard labor,
% % of whitewashing a fence, 
% manages to inveigle neighborhood boys
% into paying him for the privilege of doing this work.  While few
% businesses have attained this pinnacle of success (some subscription
% services whose main attraction is the presence and contributions
% of other members come to mind), many others have come close.
% For example, Angie's List relies entirely on user-generated material,
% but charges for access to it.  

% One of the main sources of complaints from academics is that
% publishers make huge profits out of the work of researchers,
% and not only as authors, but also as editors and reviewers.
% This of course has been true from the beginning.  It is also
% a phenomenon that is beginning to dominate our economy.
% If we look at some of the most admired and most profitable
% technology companies, such as Google and Facebook, they
% depend for their success and profits on the work of their users.
% The collective effort that social network members put into
% maintaining their 
% The key is to 

% Angie's List     That business model makes it particularly rare. There are few companies of its size that
%         are built entirely around user-generated content yet ultimately charge its contributors to
%         access just about all of it.

\section{The ``Tom Sawyer economy'' and control of chokepoints}\label{sec:toms}

A frequent complaint by researchers and librarians is that publishers profit
from the unpaid labor of scholars who write papers and referee submissions.
Yet that is an increasingly common feature of the economy, and scholarly
publishing may be viewed as a forerunner of this trend.

In the stock market, for much of 2012, the most valuable company in the world was Apple.
It is also among the enterprises most admired by the general public.  
Its dominant position is due largely to the iPhone and the iPad,
and a key reason those products were embraced enthusiastically by
the world is the multitude of apps that are available.  
Those apps are mostly produced by independent developers.
Apple boasts that these developers have earned \$6.5 billion
over four years~\cite{Streitfeld2012}.  But there are about 300,000
of these developers, so on average they have earned all of about \$20,000
on average from their creations, and this was spread over four years.
In the meantime, in just the last quarter of 2012, Apple had revenues
of \$54.5 billion, with net profits of \$13.1 billion, the bulk derived
from the iPhones and iPads.  So it is a partnership in which Apple
gets almost all the profits.  What the app developers get is
bragging rights, hopes of striking it rich, and an occasional
bonanza, to keep the game going.

The Apple app developer situation is characteristic of many of the
most visible recent high-tech success stories.  Facebook and Google
derive their value from the activities and contributions of
their users.  They only provide the basic infrastructure for
interaction.  

The value of crucial control points in extracting value was
already demonstrated by Intel and Microsoft in the PC industry.
It was a common observation that by controlling the chip
architecture and the operating system software, respectively,
this duo was earning monopoly profits, while the rest of the
industry had to be content with commodity status.  Today we
are further along the way to what we might call the ``Tom Sawyer economy.''
In Mark Twain's classic,
Tom Sawyer, condemned by Aunt Polly to a day of hard labor,
% of whitewashing a fence,
manages to inveigle neighborhood boys
into paying him for the privilege of doing this work.  While few
businesses have attained this pinnacle of success,
many have come close.
For example, Angie's List relies entirely on user-generated material,
but charges for access to it.
Among others in a similar situation is, of course, academic
publishing.

%  (some subscription
% services whose main attraction is the presence and contributions
% of other members come to mind), many others have come close.
% For example, Angie's List relies entirely on user-generated material,
% but charges for access to it.

% From this perspective, publishers making large profits from
% the much larger scholarly publishing ecosystem fits an
% increasingly common pattern.

\section{Privacy and the surreptitious spread of price discrimination}\label{sec:pricediscr}

That the University of Michigan paid \$2 million for the same unlimited
access to all Elsevier journals that the University of Montana got for less than
a quarter of the price is an excellent illustration of an extremely important,
ongoing, and so far seldom discussed development in our economy.  That is
the growth of price discrimination.

% Elsevier charging the University of Michigan \$2 million for the same unlimited
% access to all its journals that the University of Montana gets for \$0.4 million
% is an outstanding example of price discrimination in action, and an excellent
% signpost to where the entire economy is going.  

There is nothing
nefarious about price discrimination.  It is a basic technique that has been
practiced for millennia, and its theoretical foundations, which demonstrate its
frequent benefits
in stimulating economic activity, were first elucidated
by the French econo-engineers of the middle of the 19th century~\cite{EkelundH}.
Price discrimination by itself is not the ultimate goal in commerce.  Rather, 
price discrimination is
a tool for maximizing revenues and profits.  
(Some techniques, such a the
bundling represented by ``Big Deals,'' are viewed most productively as
forms of price discrimination that avoid the necessity of costly, intrusive,
and objectionable fine-scale pricing.)  
First degree price discrimination,
in which each customer is charged the maximal amount that customer is
willing and able to pay for a particular good or service, is the ideal.
However, for a long time it was regarded as mainly a theoretical construct,
almost impossible to attain (except, to some extent, in cases, such as that
of the income tax, where
governments with coercive powers obtain detailed financial
information from taxpayers under penalty of jail).  The Internet, however,
is making what was unattainable increasingly feasible.

The Internet, especially in its wireless incarnation, enables sellers
to do two things that have been difficult in the traditional brick-and-mortar
world.  Both are needed to enable first degree price discrimination,
and both require erosion of privacy.
One is to find out how much each individual purchaser is willing to pay.
% through erosion of privacy.  
The other is to control usage, which
prevents arbitrage (in which a person able to obtain a low price
resells the good or service to a person who would be charged a
higher price by the seller).  

The basic theory of the standard {\em Homo economicus}
economic model, as well as results from psychology, behavioral
economics, and especially history, lead~\cite{Odlyzko1996,Odlyzko2003} to the theory 
that the incentive to price discriminate is the driving force behind the erosion of privacy,
and to numerous predictions, such as:
\begin{itemize}
\item
there will be no limit on the extent of privacy invasion attempted by sellers
\item
the public will be very sensitive to privacy erosion
\item
the public will be extremely sensitive to price discrimination
\item
price discrimination will spread slowly and primarily in concealed forms
\item
the public will not resort to technical means for protecting privacy, but will call
for government intervention
\item
governments will play ambiguous roles, often pretending to protect privacy, but frequently
acting to facilitate price discrimination
\end{itemize}

Recent news stories provide nice illustrations of these points.
The account of how Target deduces from a customer's purchases
that she is pregnant, and starts sending her ads for infant care
items, sometimes even before her family learns of her condition,
attracted considerable public attention~\cite{Duhigg2012}.
(Rather characteristically in this area, Target stopped cooperating
with the reporter even before the story was published, surely
out of concern about such public attention.)  But the logic of
this move is clear.  It is also clear that there are incentives
to move even earlier, especially once other retailers start
competing in this arena.  Why wait to deduce the fact of
pregnancy from purchases?  If one can get hold of information
about the customer's condition, say from visits to physicians,
or from health care monitoring services, why not start then?
Or better yet, if one can monitor her voice or email communications
with her ``significant other,'' one can start a campaign even
before she gets pregnant.  

Companies detected in such practices invariably claim they
do not care about ``personally identifiable'' information,
so that, for example, they would be sending out ads for infant care
products to the class of women who might be pregnant.  This
excuse of not having individual identifications was widely
accepted, not just by industry apologists, such as the
authors of~\cite{RubinL2002} (who claimed in several places
that ``[a]dvertisers have no interest in the identity of individual customers''),
but also by more careful scholars, such as D. Solove, who opined
in 2001 (\cite{Solove2001}, p.~418)  that ``[s]ince marketers are interested in aggregate data, 
they do not care about snooping into particular people's lives.''
But such claims are ever harder to sustain in the face of data brokers
that ``are increasingly tying people's
real-life identities to their online browsing habits,'' so that
a car dealer greets a potential customer with detailed information
on what car models and accessories that customer had investigated on the Web~\cite{Valentino2012a}.
And would such information lead the dealer to offer the customer
an especially low price?

Public sensitivity to such practices is shown by the strong negative
reactions to the discovery that Orbitz was showing to Mac users
more expensive hotels than to PC users~\cite{Mattioli2012}.  This was actually a
very mild move, as even the Mac users had access to the lower
priced hotels displayed to PC users, it's just that they were
not displayed at the top of the screen.  
Still, this incident does demonstrate public concern,
and makes sellers aware they are playing with fire when
they engage in such practices.
That is why price discrimination is practiced quietly,
and mostly in mild forms so far~\cite{MikiansGEL2012,Valentino2012}.
But we can expect to see intensification of such efforts,
both as part of an arms race, and also as ways to probe
how much the public will tolerate, and to accustom the
public to such practices.

Many of the design choices in services and goods that we
observe are likely influenced by the price discrimination incentive.
For example, the push for ``The Cloud'' places people's activities
under far closer scrutiny of the operators of such clouds.
That provides more data about customer's interests and
willingness to pay, and also allows for monitoring of
usage to prevent arbitrage.
The excitement about mobile technologies is likely
also stimulated by the implicit hope that more information
and control over user's activities will translate into
large profits.  The concerns that many investors and
financial analysts express about lack of effectiveness
of mobile ads may thus miss the point, or be of only short-term
significance.  The real promise
of the always-connected mobile life for service providers
is that they might get more control over their customers.

As has often been noted, the public concern about privacy
is an unfocused one.  People complain about their loss of
privacy, but are not willing to do much about it, and
usually cannot articulate what it is they are upset about.
Framing the issue in terms of price discrimination 
brings clarity to the discussion.  On one hand there
are the well-understood benefits of this practice,
on the other hand is its assault on the sense of fairness
(which appears not to be unique to {\em Homo sapiens,}
as even some monkey have been shown to share it), and
the potential demise of the market economy, as the
demand-supply curves so beloved of economists become
irrelevant.

Academic publishers' ``Big Deals'' provide a very illuminating
example of the spread of price discrimination.
That it should happen is natural.  Technology is enabling it,
the economic incentives are there, and, after all, universities
are among the foremost practitioners of differential pricing,
with high tuition coupled with
high levels of financial aid.
So why should they not be subjected to it as well?
Until the late 1990s, pricing for scholarly journals was simple and uniform (at least
for libraries, publishers were already moving into differential
pricing by offering lower rates to individual subscribers
as well as to less-developed countries).  Now we have a 4.5 or greater differential
between state universities in the U.S., and surely much higher
across the world, where many countries do have concessionary deals.

This extensive and intensive application of price discrimination
was accomplished surreptitiously.  It is only because of the
efforts of Ted Bergstrom and his colleagues~\cite{BergstromCM}
that we have a few details, and their efforts were impeded by
publishers filing lawsuit to stop the disclosure.  (In addition
we have the indirect evidence given by the statistics of this
paper.)  One might think that Elsevier would be glad to demonstrate
that, contrary to its reputation as a predatory profit-maximizer,
it was engaging in the very socialist activity of leveling access
to information by charging the rich more than the poor.  But
such is not the case.  And it is easy to understand why, as
issues of fairness arise even in dealing with large universities.
It is easy for Harvard to appeal to its alumni to donate
so that Harvard's libraries can be the best in the world,
and provide an advantage for Harvard faculty and students.  It is
a trickier thing to ask Harvard alumni to donate to their
{\em alma mater} just so it can pay ten times as much
as some unknown school in the boondocks for the same degree
of access to Elsevier journals.

Another noteworthy point about the spread of ``Big Deals'' is
how slow it was.  Why did the developments visible in Fig.~\ref{fig:serials} 
not occur a decade earlier?  Part of the reason is surely that
publishers have been slow to learn.  They spent much of the
late 1990s playing around with schemes to charge individuals
for each access.  These experiments, such as the PEAK one run
by Elsevier, were costly and annoying to the scholars who
got involved~\cite{LougeeM2008}.  They should have been seen
as doomed right from the beginning, but the industry had to
learn this lesson for itself.  And even after they realized
that ``Big Deals'' was the best way to proceed, it took a while
for it to be widely accepted.  Librarians may not be as conservative
in their ways as academics, but still the terms of the contracts
had to be hammered out, the budgeting and contracting procedures changed, etc.
Note that the sensitivity to price discrimination in dealing
with academia is not as severe as in dealing with the public.
In general, differential pricing is far more common in
business to business transactions.  
% Furthermore, universities themselves form one of the outstanding and best documented examples
% of price discrimination, with high tuition coupled with high levels of financial aid.
This slow spread of price discrimination in academic publishing
likely foreshadows its slow spread in consumer pricing as well.

One reason the differential pricing practices of the ``Big Deals''
could be implemented effectively was the gain in information by
publishers about usage of their products.  In the old print world,
they knew very little about it.  The librarians knew more, but
even the librarians had limited knowledge, not much more than the frequency of
circulation of print volumes.  With the move to online distribution,
much more precise measures became available, and now the publishers
have the upper hand.  While they do share some information with
libraries, they can aggregate and compare data from different
institutions, and of course they have information about library
and entire university budgets.  Thus this offers another example
of how loss of privacy facilitates price discrimination.

Some other publisher practices are also what we find elsewhere.
Prices are always presented as discounted from some notional
list price, and never as surcharges imposed on rich institutions.
From the standpoint of the standard economic model, it does
not matter which method is used.  But {\em Homo sapiens} is
not {\em Homo economicus,} and how prices are presented matters
a lot.  That's why universities practice price discrimination
by having high tuition and offering frequent and large discounts
(called scholarships and the like).
And that is why we are moving to a world in which Starbucks will likely
raise the standard price of their Caramel Frappuccino to \$9.95,
and never discount it below \$8 for somebody who is rich and
a Caramel Frappuccino addict, while somebody who can't tell
the difference between a Caramel Frappuccino from Starbucks
and a regular cup of Macdonald's coffee will get discounts lowering
the effective price to \$2.  (The mobile world, where the
nature of the discount can be obfuscated due to location
and other circumstances, will likely facilitate such developments.)

The final major lesson from ``"Big Deals'' is that 
however hard the public may dislike price discrimination, it
is hard to stop.  
Sellers will have strong incentives to employ it, especially as competition
intensifies.  With their resources, and incentives to
cooperate, consumers are unlikely to be able to protect themselves.
Further, even if they could, it is unlikely they will bother.
Historical precedents suggest they will turn to the government
to step in and regulate objectionable practices~\cite{Odlyzko1996,Odlyzko2003}.
However, governments have often played ambiguous roles, paying
lip service to public demands, but facilitating price discrimination.
Some differential pricing can be criticized for
targeting the poor (or, more frequently, the ignorant, who are often, but
not always, poor)~\cite{Valentino2012}.
But very often it can be shown to have positive social effects.
Such is the case with progressive taxation, with college
tuition practices, and with prescription drug price differences
between countries~\cite{Lichtenberg2010}.
The murkiness of many of these deals does leave ground for
criticism.  (For example, one index shows drug prices in
Mexico as 24\% higher than in the U.S., which is high on 
the list, but not the highest.)  But there is a ``highly
significant positive correlation between per capita income
and the drug price index,'' so on balance it is easy to argue
they are good for the world as a whole.
The ``Big Deal'' practices also appear to have enhanced
social welfare.  One can argue that the world would have
been better had libraries and researchers moved to squeeze
out publishers and move Open Access forward faster.
But both groups have been slow to act, while publishers
are providing visibly better service through ``Big Deals,''
as is seen in Fig.~\ref{fig:serials}.  Hence why should
even well-intentioned regulators step in?
Chances are that similar arguments will prevail in many
other situations, and that in spite of public opposition,
price discrimination will spread.

\section{Conclusions}\label{sec:conclusions}

The most important developments in scholarly publishing are taking
place quietly, with essentially no public discussion.  
The traditional roles of the two key intermediaries, libraries and
publishers, are shrinking.  New opportunities are opening up,
however, which forces those two groups into new roles, and brings
in new agents (such as Google).  Who will perform what functions
is open to competition, and at this point the publishers are
doing better than libraries.  Whether this will continue,
and what the role of new agents that are competing for
these functions will be is hard to predict.  However,
publishers have provided much better access to scholarly
information through the often criticized ``Big Deal'' packages.

The evolution of the academic information market suggests
that the general economy will continue to evolve in ways
contrary to common predictions.  Opaque markets and
price discrimination are likely to play an increasing role.

% that changing
% technologies are creating are open to competition making possible are open to competition by
% new agents (such as Google).  Who will perform what functions
% is open to competition, and at this point the publishers are
% doing better than libraries.  Whether this will continue,
% and what the role of new agents that are competing for
% these functions will be is hard to predict.
% Inertia of the scholarly publishing system is immense.

% \clearpage

\begingroup
\parindent 0pt
\def\enotesize{\normalsize}
\theendnotes
\endgroup

% \clearpage

\section*{Acknowledgments}

Thanks are due to Claudio Aspesi, Ted Bergstrom, Kris Fowler, Paul Ginsparg, Stevan Harnad,
and Jonathan Smith
for providing helpful information or comments.

% The many individuals and institutions that assisted in the

\end{document}